\documentclass[aps,prd,english,superscriptaddress,11pt,notitlepage]{revtex4}
	\usepackage[colorlinks=true, a4paper=true, pdfstartview=FitV,
linkcolor=blue, citecolor=blue, urlcolor=blue]{hyperref}

\usepackage{amsmath}
\usepackage{amssymb}
\usepackage{graphicx}
\usepackage{hhline}
\usepackage{textcomp}
\makeatletter
\usepackage{babel}
\newcommand{\bea}{\begin{eqnarray}}
\newcommand{\eea}{\end{eqnarray}}

\newcommand{\e}{\epsilon}

\newcommand{\Tr}{{\rm Tr}}

\newcommand{\be}{\begin{equation}}
\newcommand{\ee}{\end{equation}}

\usepackage[active]{srcltx}
\begin{document}


\title{Carbon-12 in the generalized Skyrme model}

\author{Christoph Adam}
\affiliation{Departamento de F\'isica de Part\'iculas, Universidad de Santiago de Compostela and 
Instituto Galego de F\'isica de Altas Enerxias (IGFAE) Santiago de Compostela, E-15782, Spain}

\author{Carlos Naya}%

\affiliation{
 Institute of Theoretical Physics,  Jagiellonian University, Lojasiewicza 11, 30-348 Krak\'{o}w, Poland
}

\affiliation{
Department of Physics, Stockholm University, AlbaNova University Center, 106 91, Stockholm, Sweden
}

\author{Andrzej Wereszczy\'{n}ski}

\affiliation{
 Institute of Theoretical Physics,  Jagiellonian University, Lojasiewicza 11, 30-348 Krak\'{o}w, Poland
}
 \affiliation{Department of Applied Mathematics, University of Salamanca, Casas del Parque 2 and Institute of Fundamental Physics and Mathematics,
University of Salamanca, Plaza de la Merced 1, 37008 - Salamanca, Spain
}

 \affiliation{
 International Institute for Sustainability with Knotted Chiral Meta Matter (WPI-SKCM2), Hiroshima University, Higashi-Hiroshima, Hiroshima 739-8526, Japan
 }

\begin{abstract}
We study properties of the $^{12}$C nucleus within the generalized Skyrme model where, in addition to the standard massive Skyrme model, the sextic term and the pionic potential squared are included. We find that the model continues to accurately describe the rotational bands of the $^{12}$C nucleus. In addition, at variance with the case of the standard Skyrme model, it provides the correct energy ordering of the classical solutions which correspond to the ground state and the Hoyle state, respectively.
\end{abstract}
\maketitle

\section{Motivation}
The Skyrme model \cite{Skyrme}, in principle, allows for a unified field theoretical description of baryonic matter at all scales - from single baryons to infinite nuclear matter \cite{HS-book, MR-book, NM-book}. Specifically, the description of atomic nuclei in the Skyrme model framework is based on Skyrmions i.e., classical topological soliton solutions of the model, where their topological charge $B$ can be rigorously identified with the baryon number. Further quantum numbers of nuclei, like spin and isospin, are described by the semiclassical quantization of Skyrmions, i.e., the quantization of the lightest collective degrees of freedom of the classical solutions, like the zero modes (the classical symmetries) \cite{ANW} or the vibrational modes (massive deformations) \cite{H-1}. 

Until about a decade ago, most calculations within this framework were restricted to the standard Skyrme model (to be defined below) and to the semiclassical quantization of the zero modes, the so-called  rigid rotor quantization.
Amongst the successes of this approach, let us mention the very good description of the excitation bands of several light atomic nuclei \cite{Manko2007, Battye2009}.
Nevertheless, for other nuclei, particularly the ones with an odd baryon number $B$, the nuclear spectra are not reproduced correctly, indicating that probably both generalized versions of the Skyrme model and a more refined quantization procedure are required. 

Indeed, the implementation of the vibrational quantization \cite{H-1}, that is, the inclusion of the massive modes, led to spectacular results for the nuclei $^7$Li \cite{H-1} and $^{16}$O \cite{H-2,H-3}. Also $^{12}$C found an elegant and quite accurate description within the Skyrme model. The Carbon-12 nucleus is known to exist in two spin zero states:  the ground state and the slightly heavier Hoyle state \cite{Hoyle}. Both give rise to excitation bands with different slopes.
 The Skyrme model provides a very elegant explanation of these facts \cite{Lau2014}. For the $B=12$ Skyrmion, there are two geometrically distinct classical solutions related to a different organization of the $B=4$ ($\alpha$ particle) substructures \cite{Battye2007,Rawlinson2018}. The first one has the $D_{3h}$ symmetry, where three $B=4$ Skyrmions are arranged into a triangular shape. The second one, with the $D_{4h}$ symmetry, is a straight chain of $B=4$ Skyrmions. Different shapes of the classical solutions result in unequal spin inertial tensors and, as a consequence, lead to different slopes of the excitation bands, which agree with the experimentally measured ones to a very good precision. In fact, this is the Skyrme model realization of the $\alpha$-particle clustering model of light nuclei.

There has been, however, one issue with the Skyrme model computation. The triangular Skyrmion has slightly higher energy than the chain solution. Thus, while the slopes were correct, the energy ordering of the states did not agree with the experimental data. 

In the present paper, we show that a certain generalization of the Skyrme model gives the correct order of the ground and Hoyle states, while keeping simultaneously the slopes of the excitation bands close to the experimental values. This is achieved by the inclusion of the so-called BPS part \cite{BPS}, which combines a term which is sextic in first derivatives \cite{sextic} and a new potential term \cite{BG-0, BG-1, BG-2}. 

\section{Landscape of classical Skyrmions}

The static energy functional of the generalized Skyrme model which we consider in the present paper reads
\bea
E&=&E_{Sk} + E_{BPS} =  \nonumber\\
&=& \int \left( -\frac{f_\pi^2}{4} \mbox{Tr} \left( R_iR_i\right) -\frac{1}{32 e^2} \mbox{Tr} \left[R_i,R_j\right]^2-
\frac{1}{8} m_\pi^2 f_\pi^2 \mbox{Tr} \left( 1-U \right) \right)d^3x \nonumber \\
&+& \int \left(  4\pi^4 \lambda^2 \mathcal{B}^2 +   \frac{M^2_2}{4} \left[ \mbox{Tr} \left(1-U\right) \right]^2 \right)d^3x ,
\eea
with $f_\pi$  the pion decay constant, $m_\pi$ the pions mass, and $e$, $\lambda$ and $M_2$ some coupling parameters. Besides, $R_i=\partial_i U U^\dagger$ is the right invariant current of the $SU(2)$ valued matrix Skyrme field $U(\boldsymbol{x})$. In order to have finite energy solutions the field must take a constant value at infinity, in particular, $U(|\boldsymbol x| \rightarrow \infty) = 1$, allowing us to compactify the base space $\mathbb R^3$ into $S^3$. Hence,
\be \label{Bden}
\mathcal{B}=- \frac{1}{24\pi^2} \epsilon_{ijk} \mbox{Tr} \left( R_iR_jR_k \right)
\ee
is the baryon charge density whose volume integral gives the topological baryon charge $B$. 

In comparison with the original Skyrme proposal, which we call the massive Skyrme model $E_{Sk}$, it has an additional contribution from the so-called BPS Skyrme model $E_{BPS}$ \cite{BPS}, which contains two additional terms. Namely, the baryonic density squared term (referred to as the {\it sextic} term \cite{sextic}) and the pion potential squared term \cite{BG-0, BG-1, BG-2}. It is important to note that this new non-derivative part does not influence the small field expansion and, therefore, it does not change the pion mass, $m_\pi$. 

The original motivation to include the BPS model is related to the fact that the model $E_{Sk}$ leads to rather large classical binding energies, which are much larger than the binding energies of physical nuclei. The BPS model alone has zero classical binding energies \cite{BPS}, therefore its inclusion should ameliorate this problem. Later it was found that a new potential term (the fourth power of the pion mass potential) serves to reduce binding energies by itself \cite{DH}, \cite{MS}. This observation was generalized to other types of potential (e.g., the square of the pion mass potential) with or without the presence of the sextic term \cite{BG-0, BG-1, BG-2}. 
The sextic term is, however, indispensable for a realistic description of nuclear matter at high densities, because without it the resulting nuclear matter equation of state turns out much too soft \cite{Sk-EOS}.

For convenience, we can define the energy and length units, allowing us to write a dimensionless static energy functional depending only on three coupling constants instead of five. We will choose to keep the coefficients of the kinetic term and the pion mass potential fixed. To this purpose, we rewrite the $e$ coupling constant as follows, $e^2 \rightarrow e^2/c_4$ and introduce the usual choice of Skyrme units
\be
l= \frac{2}{ef_\pi},  \;\;\; E_0=\frac{f_\pi}{4e}.
\ee
Hence, the static energy functional of the generalized Skyrme model reads now
\bea \label{Energy}
E&=& \frac{1}{12\pi^2} \int \left( -\frac{1}{2} \mbox{Tr} \left( R_iR_i\right) -\frac{c_4}{16} \mbox{Tr} \left[R_i,R_j\right]^2+
m^2\mbox{Tr} \left( 1-U \right) \right)d^3x \nonumber \\
&+& \frac{1}{12\pi^2} \int \left(  4\pi^4 c_6 \mathcal{B}^2 +   \frac{m^2_2}{4} \left[ \mbox{Tr} \left(1-U\right) \right]^2 \right)d^3x ,
\eea
where the mass parameter $m$ is related to the physical pion mass $m_\pi$ as 
\be \label{mass}
m=\frac{2m_\pi}{ef_\pi},
\ee
while the relation of the new coupling constants $c_6$ and $m_2$ with the previous ones is not important. Furthermore, we have normalized the total energy by a factor $1/12\pi^2$ which, together with the Faddeev-Skyrme BPS bound \cite{F}, implies that $E_{Sk} \geq \sqrt{c_4} |B|$. The model contains now three tunable parameters $c_4, c_6$ and $m_2$, which are the couplings of the quartic Skyrme term, the sextic term and the additional potential, respectivley The pion mass parameter is assumed to be $m= 0.526$, which provides the physical mass of the pions if the standard calibration of the model $E_{Sk}$ is chosen. 

As we already noticed, in the massive Skyrme model (where $c_6=m_2=0$), the $B=12$ chain Skyrmion has lower energy than the triangle solution. In our numerical analysis we varied $c_4,c_6$ and $m_2$ in a search for a region where:
\begin{enumerate}
\item[{\it i)}] the $D_{3h}$ and $D_{4h}$ Skyrmions, with a well pronounced $\alpha$-particle clustering structure, are stable critical point of the energy functional and 
\item[{\it ii)}] the triangular solution with $D_{3h}$ symmetry has lower energy that the chain solution.
\end{enumerate}

In this process, we have minimised the energy functional for different values of the coupling constants and initial conditions using the same numerical scheme as in \cite{Naya2018}. In summary, this consists in evolving the system through second-order-in-time equations derived from a Lagrangian with the static contribution given by Eq. (\ref{Energy}) in a cubic grid with boundary condition $U=1$. Spatial derivatives are approximated by fourth-order finite differences and the time evolution is carried out by a fourth-order Runge-Kutta method. For the system to flow to the minimal configuration, the motion is immediately frozen by setting the time derivatives to zero when the static energy increases before continuing the further evolution. During the general scanning of the parameter space, we have considered a lattice of $100^3$ points with spacing $\Delta x = 0.15$ and  time step $\Delta t = 0.08$. 

After an exhaustive sweeping of the coupling constants, we have found an optimal set of values where the coefficients of the quartic and sextic contributions are equal to one, $c_4 = c_6 = 1$, while $m_2 \geq 0.5$. For smaller values of $m_2$ we do not find the $D_{3h}$ soliton, even if we started with a suitable initial configuration. In this case we arrive at the configuration with $C_{3v}$ symmetry described in \cite{Battye2006}. This seems to indicate that for $m_2 \lesssim 0.5$ the desired Skyrmion describing the carbon-12 ground state is a saddle point solution. For the same set of parameters the $D_{4h}$ states become basically unchanged. These solutions are chain clusters with well visible $\alpha$ particles.

\begin{figure}
\begin{center}
\vspace*{-.8cm}
\includegraphics[width=0.49\textwidth]{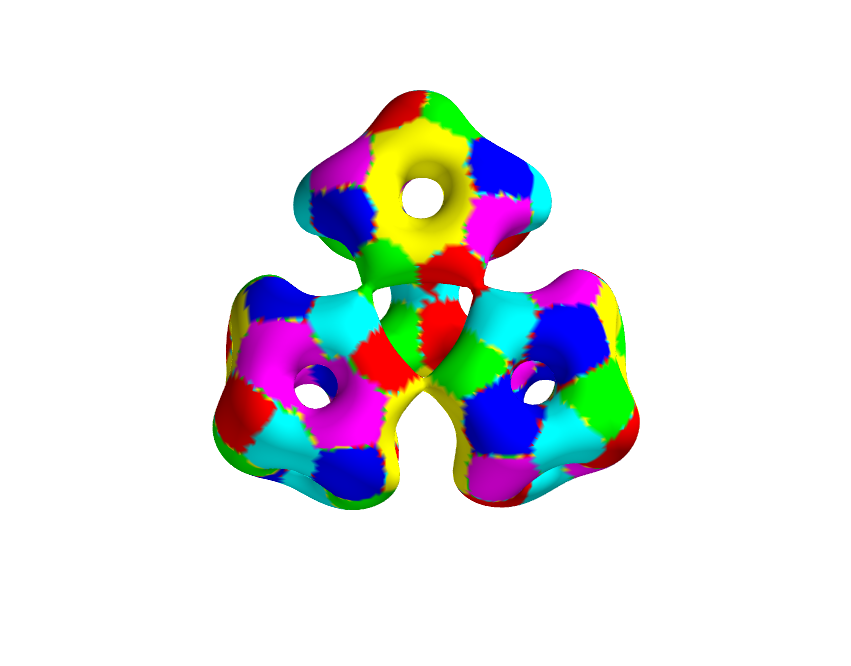}
\includegraphics[width=0.49\textwidth]{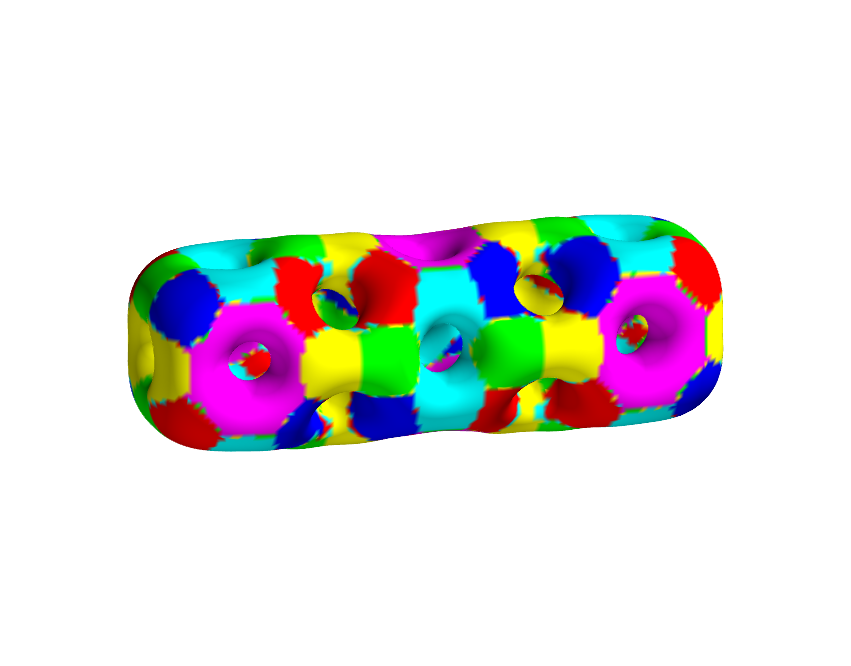}
\vspace*{-.8cm}
\end{center}
\caption{Baryon density isosurfaces for Skyrmions with baryon number $B=12$ corresponding to the $D_{3h}$-symmetric ground state (left) and the $D_{4 h}$-symmetric Hoyle state (right).} \label{Fig_GS&Hoyle}
\end{figure}

\begin{table} 
\begin{center}
\begin{tabular}{ccccccccccc} 
\hline \hline
 & & $\quad E$ & & $E/B$ & & $b$ & & $V_{11}$ & & $V_{33}$ \\
 \hline
 Ground state & & 17.3418 & & 1.4452 & & 0.99993 & & 12540.6 & & 18363 \\
 \hline
  Hoyle state & & 17.3553 & & 1.4464 & & 0.99994  & & 26399.3 & & 5462.2 \\
\hline \hline
\end{tabular}
\caption{Numerical results for the $D_{3h}$ and $D_{4h}$ symmetric Skyrmions describing the ground and Hoyle states, respectively. The energy, $E$, and energy per baryon number, $E/B$, are given in Skyrme units normalised by the Skyrme-Fadeev bound, while $b = B/12$ is the corresponding baryon number unit. Similarly, Skyrme units have been used for the spin moment of inertia tensor eigenvalues.}
\label{Tab_states}
\end{center}
\end{table}

In the region of parameters that lead to the desired landscape of classical solutions we increased our accuracy, with a lattice of $148^3$ points, $\Delta x = 0.1$ and $\Delta t = 0.05$. In fact, for $m_2=0.5$ we found the optimal situation where the mass splitting between the ground and Hoyle states from the generalized Skyrme model is about $0.08\%$. Hence for this generalized Skyrme model we not only get the correct ordering of the ground and Hoyle states but its splitting is comparable to the experimental value of $0.07\%$. The optimal solution is plotted in Fig. \ref{Fig_GS&Hoyle}.  In the figure the baryon density isosurfaces are shown with a colouring convention based on the parametrisation of the Skyrme field in terms of a triplet of pionic fields $\boldsymbol \pi = (\pi_1,\pi_2,\pi_3)$ and an auxiliary $\sigma$ field, $U = \sigma + i \boldsymbol{\pi} \cdot \boldsymbol{\tau}$, such that $\sigma^2 + \boldsymbol{\pi}^2 = 1$, with $\boldsymbol \tau$ the triplet of Pauli matrices. Then, regions with a dominant positive (negative) $\pi_1$, $\pi_2$ or $\pi_3$ fields are coloured in magenta (green), yellow (cyan) or red (blue), respectively.
The classical energy, $E$, energy per baryon number, $E/B$, and baryon number unit $b = B/12$, i.e., the numerically-obtained baryon number divided by 12, are shown in Table \ref{Tab_states}.

\section{Spin quantization and Coulomb energy}

In order to have a complete description of the carbon-12 states, two additional contributions to the energy should be taken into account: the spin excitations, giving rise to rotational bands, and the Coulomb energy. To implement the former, one has to quantize the Skyrmions. This is usually done through a semi-classical quantization of the rotational and iso-rotational degrees of freedom which give rise to the spin and isospin quantum numbers \cite{ANW, Manko2007}. In this procedure one introduces $B$ rotational and $A$ isorotational transfomations of the classical Skyrmion $U(\boldsymbol{x}) \to A U(R_B\boldsymbol{x}) A^\dagger$  and promotes them to time-dependent collective coordinates. Here $A,B \in SU(2)$ and $R_B=\frac{1}{2} \mbox{Tr} (\tau_iB\tau_JB^\dagger)$, where $\tau_i$ are Pauli matrices. Then we arrive at a (mechanical) Hamiltonian which is quantized in the canonical way. In fact, the resulting Hamiltonian is similar to a rigid rotor one and is given in terms of the corresponding moment of inertia tensors of spin and isospin together with the spin and isospin operators, for a review see \cite{NM-book}. 

Since the carbon-12 has zero isospin, we only need to consider the spin moment of inertia tensor, $V_{ij}$, which for the generalized Skyrme model reads (remember, $c_4=c_6=1$ for the optimal choice)
\be \label{Vij}
V_{ij} = -\epsilon_{ikl} \epsilon_{jmn} \int x_k x_m \left[ \Tr \left(R_l R_n +  \frac{c_4}{4} [R_p, R_l] [R_p, R_n] \right) - c_6 \Tr(\e_{pqr} R_l R_q R_r) \Tr(\e_{pst} R_n R_s R_t) \right] d^3 x.
\ee
Due to the symmetry of the configurations, the tensor $V_{ij}$ is that of a symmetric top, that is to say, diagonal with two equal eigenvalues, $V_{11} = V_{22}$, different from the third, $V_{33}$. Hence, the corresponding quantum contribution takes the form
\be \label{Erot}
E_{\rm rot} = \frac{1}{2 V_{11}} J(J+1) + \left( \frac{1}{2 V_{33}} - \frac{1}{2 V_{11}} \right) k^2,
\ee
where $J$ corresponds to the spin quantum number and $k$ is the third component of the spin in the body-fixed frame. The obtained eigenvalues of the spin moment of inertia tensor for our solutions of interest are also displayed in Table \ref{Tab_states}.

It is very well known that the discrete symmetries of the solitons impose some constraints restricting the allowed quantum states. These are the Finkelstein-Rubistein constraints \cite{Finkelstein1968} and have been already discussed in the literature for both configurations \cite{Battye2009,Lau2014}. In particular, the ground state symmetry $D_{3h}$ allows for states with $k$ being zero or a multiple of 3, while for the Hoyle solution, states with $k=0$ and 4 are predicted. Since these results are entirely due to the symmetry of the classical Skyrmions, they remain unchanged in the case of our solutions obtained in the generalized Skyrme model. As pointed out in \cite{Lau2014}, this offers a successful description of the rotational bands of both states of carbon-12 within the Skyrme model, leading also to the prediction of new states. For instance, it naturally supports the $0^+$, $2^+$ and $4^+$ states of the ground state band with excitation energies 0, 4.4 and 14.1 MeV, together with those of the Hoyle band having excitation energies 7.65, 9.8 and 13.3 MeV (for the states we have used the common notation $J^P$, where $J$ and $P$ correspond to the total spin and parity, respectively). 

\vspace*{0.2cm}

Finally we consider the Coulomb energy, which can be expressed in terms of the electric charge density of the soliton by its generalization for volume charge densities,
\be \label{EC}
E_{\rm C} = \frac{1}{2 \varepsilon_0} \int \int d^3 r d^3 r' \frac{\rho (\boldsymbol r) \rho(\boldsymbol{r'})}{4 \pi | \boldsymbol r - \boldsymbol{r'}|},
\ee
where $\varepsilon_0 = \frac{1}{\rm e} \, 8.8542 \cdot 10^{-21} \, {\rm MeV}^{-1} \, {\rm fm}^{-1}$ is the permittivity of free space with ${\rm e} = 1.60218 \cdot 10^{-19}$ C. Since for carbon-12 the isospin vanishes, its electric charge density is just half the baryon charge density as given in Eq. (\ref{Bden}).

This contribution to the energy has been previously computed both within the massive Skyrme model \cite{Ma2019} and its BPS counterpart \cite{Adam2013a,Adam2013b}. Here we will follow the same strategy which is based on the multipole expansion of the Coulomb potential \cite{Carlson1963}
\be
\frac{1}{4 \pi |\boldsymbol r - \boldsymbol{r'}|} = \sum_{l=0}^\infty \sum_{m=-l}^{l} \frac{1}{2 l+1} \frac{r_<^l}{r_>^{l+1}} Y^*_{lm}(\theta',\phi') Y_{lm} (\theta, \phi),
\ee
with $r_< = {\rm min} (r, r')$, $r_> = {\rm max} (r, r')$ and $Y_{lm} (\theta, \phi)$ the usual spherical harmonics. Then, by considering the expansion of the charge density in spherical harmonics
\be
\rho(\boldsymbol r) = \sum_{l, m} \rho_{lm} (r) Y^*_{lm}(\theta, \phi),
\ee
and defining the quantities
\be
Q_{lm} (r) = \int_0^r dr' r'^{l+2} \rho_{lm} (r'), \qquad \qquad U_{lm} = \frac{1}{2 \varepsilon_0} \int_0^\infty dr r^{-2 l -2} |Q_{lm} (r)|^2, 
\ee
the Coulomb energy can be reduced to
\be
E_{\rm C} = \sum_{l=0}^\infty \sum_{m=-l}^l U_{lm}.
\ee
This is a convergent series in the multipole index $l$ which we truncate to a value high enough ($l = 30$ in our case) such that the change in the Coulomb energy is less than $10^{-4}$  MeV.

\section{Results}
To discuss the physical content of our findings we have to go from Skyrme units to physical units. This is straightforward for the classical energy of the Skyrmion given by the energy functional Eq. (\ref{Energy}). It simply requires multiplication by the energy scale $f_\pi/(4 e)$.  The next contribution to consider is the spin quantum correction. In this case, we need to bring the physical units to the moment of inertia tensor, Eq. (\ref{Vij}), multiplying it by the energy scale and the length scale squared. As a result, the spin excitations gain a global factor $e^3 f_\pi$. Finally, we can see from Eq. (\ref{EC}) that to convert the Coulomb energy to physical units we need to introduce a factor equal to the inverse of the length scale, namely, $e f_\pi/2$.

Hence, after conversion to physical units, the total energy of the semiclassically quantized Skyrmions with Coulomb interaction included will read 
\be
E_{\rm ph} = \frac{f_\pi}{4 e} \, E + e^3 f_\pi \, E_{\rm rot} + \frac{e f_\pi}{2} \, E_{\rm C}.
\ee

Now, we have to calibrate our model, that is to say,  find suitable values for the constants $f_\pi$ and $e$. For this purpose, we will fit the ground state energy, $E_{\rm GS} = 11177.9 \; {\rm MeV}$, and the ground state rotational band  slope to the physical values (with excitation energies 0, 4.4 and 14.1 MeV, respectively).  This is equivalent to impose $e^3 f_\pi = 17740.0$ MeV and to set the energy per baryon number of our corresponding Skyrmion to one atomic mass unit -- amu, ($1 \; {\rm amu} = 931.494 \; {\rm MeV}$). Hence, our calibration reads
\be \label{Calibration}
e = 5.34383, \qquad \qquad f_\pi = 116.256 \; {\rm MeV}.
\ee
This also implies a value of the pion mass $m_\pi = 163 \; {\rm MeV}$, which is basically equal to the best value obtained in \cite{Lau2014} (162 MeV) and  which is quite close to the experimental value of 138 MeV. 

\begin{table} 
\begin{center}
\begin{tabular}{ccccccccc} 
\hline \hline
 & & $\quad E$ & & $E_{\rm rot} \; (2^+)$ & & $E_{\rm rot} \; (4^+)$ & & $E_{\rm C}$  \\
 \hline
 Ground state & & 11171.34 & & 4.24 & & 14.15 & & 6.56  \\
 \hline
  Hoyle state & & 11180.04 & & 2.02 & & 6.72 & & 6.02 \\
\hline \hline
\end{tabular}
\caption{Soliton classical energy, $E$, rotational, $E_{\rm rot}$, and Coulomb energy, $E_{\rm C}$, contributions to the ground and Hoyle states of the carbon-12 in physical units (MeV) after calibration, Eq. (\ref{Calibration}). In the case of the rotational energy we omit the $0^+$ states with vanishing contribution.}
\label{Tab_energies}
\end{center}
\end{table}

Taking these values we find  the different contributions to the energy of the ground and Hoyle states in physical units as shown in Table \ref{Tab_energies}. In particular, the energy of Hoyle state is
\be
E_{\rm Hoyle} = 11186.1 \; {\rm MeV},
\ee
which leads to a mass splitting at the level about $0.07 \%$, matching the value found in nature. In this way, the inclusion of additional terms to the Skyrme Lagrangian seems to be unavoidable in order to correct this issue of a wrong ordering in energy of the two configurations that is found in the Skyrme model, both in its standard version and in some extensions such as the coupling to rho mesons \cite{Naya2018}.

\begin{figure}
\begin{center}
\includegraphics[width=0.6\textwidth]{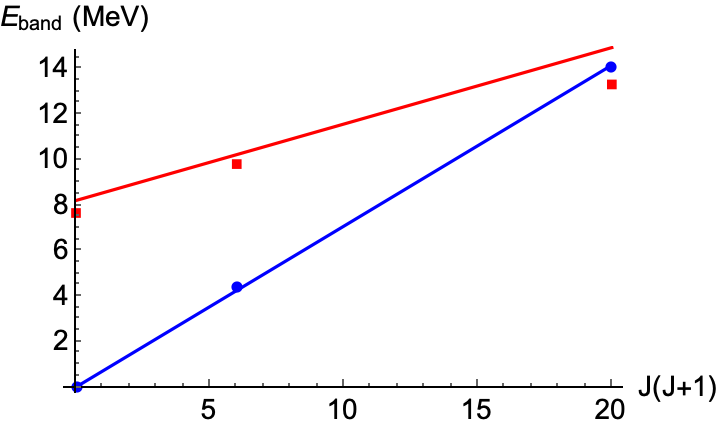}
\end{center}
\caption{Rotational bands for the ground (blue line) and Hoyle (red line) states obtained from the generalised Skyrme model. They show the excitation energy with respect to the ground state. Blue circles and red squares indicate the experimentally determined states $0^+$, $2^+$ and $4^+$ of the ground and Hoyle states, respectively.} \label{Fig_bands}
\end{figure}

Importantly, the inclusion of the new terms in the Lagrangian does not spoil the slope of the Hoyle band. It is still close to the physical value, see in Fig. \ref{Fig_bands}. Indeed, the ratio between the slopes of the ground and the Hoyle bands is 2.11 in our model, so it is a little bit below the experimentally estimated value of 2.45. In the massive Skyrme model this ratio was slightly above the experimental value, 2.52.

Finally, we can also study the root mean square matter radius, $\langle r^2 \rangle ^{1/2}$, defined as
\be
\langle r^2 \rangle ^{1/2} = \left(\frac{\int \mathcal{E}(\boldsymbol x) r^2 d^3 x}{\int \mathcal{E}(\boldsymbol x) d^3 x} \right)^{1/2},
\ee
with $\mathcal{E}(\boldsymbol x)$ the static energy density given by the integrand of Eq. (\ref{Energy}). We find the values 2.14 fm and 2.44 fm for the ground and Hoyle states, respectively. These results are slightly under the experimental value of 2.43 fm for the ground state \cite{Tanihata1985} and an inferred 2.89 fm for the Hoyle state \cite{Danilov2009}. We should note that our values here depend on the calibration of the model; however, their ratio does not. We obtain then a ratio of 1.14 to be compared with 1.19, resulting in a good agreement.

\section{Summary}

In this work, we have shown that the inclusion of two additional terms to the massive Skyrme model, the sextic term and the pion mass potential squared, fixes one main issue of the description of $^{12}$C nuclei in terms of Skyrmions, namely the correct energy ordering of the $D_{3h}$ symmetric triangle and the $D_{4h}$ chain, already for the classical soliton solutions. These solutions were interpreted as corresponding to the ground and the Hoyle states respectively. However, in the massive Skyrme model the triangle solution has slightly higher energy than the chain solution which, obviously, does not correspond to the correct physical order. The inclusion of the two further terms leads to the right ordering. Importantly, this extension of the Skyrme theory does not spoil the $\alpha$-particle clustering in the Skyrmions and has only a small impact on the spin moment of inertia. Therefore, we still get essentially the correct slopes of the excitation spectra built on top of both zero spin states, besides a remarkable agreement for their root mean square matter radius ratio. 

In a wider perspective, the present work belongs to the class of investigations which try to determine the most important extensions and generalizations of the Skyrme model which are necessary for a quantitatively precise description of physical nuclei and nuclear matter. The original Skyrme model is a fascinating proposal which incorporates many nontrivial properties of low-energy QCD in a natural way, but its quantitative precision is typically at the 30\% level, and some results, like the shapes or excitation spectra of some nuclei,  are simply not correct. Therefore, possible generalizations \cite{sextic,marleau} were considered already soon after the interest in the Skyrme model resumed \cite{ANW}, but the systematic study of generalizations of the Skyrme model only started about one decade ago, partly motivated by the more powerful numerical resources which are available nowadays.
In this context, we just want to point out that
the two new terms we included are very natural in the solitonic Skyrme model framework. The sextic term is unavoidable in the high density regime, where it provides the most important contribution, leading to a sufficiently stiff equation of state which, e.g.,  supports neutron stars with sufficiently high masses, in agreement with observational data \cite{Naya-NS,SK-NS-1,SK-NS-2,SK-NS-3}. Furthermore, both this term and the additional potential, which together form the BPS Skyrme model, provide a way to reduce the binding energies of Skyrmions, which are excessively large when compared with physical nuclei.


\section*{Acknowledgements}
The authors thank David Foster for his work at the initial stage of the project. CN thanks Miguel Huidobro for discussions and also running some Skyrmions at the initial stage of the parameter space exploration, serving as a double check. CA acknowledges financial support from the Ministry of Education, Culture, and Sports, Spain (Grant No. PID2020-119632GB-I00), the Xunta de Galicia (Grant No. INCITE09.296.035PR and Centro singular de investigacion de Galicia accreditation 2019- 2022), the Spanish Consolider-Ingenio 2010 Programme CPAN (CSD2007-00042), and the European Union ERDF. CN and AW were supported by the Polish National Science Centre (NCN 2020/39/B/ST2/01553). This research utilized the Sunrise HPC facility supported by the Technical Division at the Department of Physics, Stockholm University.

\end{document}